\documentclass{caosp302}

\usepackage{graphicx}

\articleNo{123}
\pubyear{2007}
\volume{1}
\volnumber{1}
\firstpage{1}
\received{???}
\accepted{???}

\newcommand{\kms}{km\,s$^{-1}$}
\newcommand{\vs}{$v \sin i$}
\newcommand{\teff}{$T_{\rm eff}$}
\newcommand{\lgg}{$\log\,{g}$}

\begin{document}

%
\hauthor{C.P.\,Folsom, G.A.\,Wade, {\it et al.}}

\title{Magnetic, Chemical and Rotational Properties of the Herbig Ae/Be Binary System HD 72106}

%
\author{C.P.\,Folsom \inst{1,} \inst{2,} \inst{3}
\and 
  G.A.\,Wade \inst{2}
\and 
  O.\,Kochukhov \inst{4}
\and
  E.\,Alecian \inst{2}
\and
  C.\,Catala \inst{5}
\and
  S.\,Bagnulo \inst{1}
\and
  J.D.\,Landstreet \inst{6}
\and
  D.A.\,Hanes \inst{3}
}

%
\institute{
Armagh Observatory, College Hill, Armagh, BT61 9DG, Northern Ireland,U.K. 
\and 
Department of Physics, Royal Military College of Canada, \\ P.O. Box 17000 Station `Forces' Kingston, Ontario, Canada K7K 7B4
\and 
Department of Physics, Engineering Physics \& Astronomy, Queen's University, Kingston, Ontario, Canada, K7L 3N6
\and
Department of Astronomy and Space Physics, Uppsala University, \\751 20 Uppsala, Sweden
\and
Obs.\,de Paris, LESIA, 5 place Jules Janssen, 92195 Meudon Cedex, France
\and
Physics \& Astronomy Department, The University of Western Ontario, London, Ontario, Canada N6A 3K7
          }

\date{December 1, 2007}

\maketitle

\begin{abstract}
Recently, strong, globally-ordered magnetic fields have been detected in some 
Herbig Ae and Be (HAeBe) stars, suggesting a possible evolutionary connection 
to main sequence magnetic chemically peculiar Ap and Bp stars. 
We have undertaken a detailed study of the binary system HD 72106, which contains 
a B9 magnetic primary and a HAeBe secondary, using the 
ESPaDOnS spectropolarimeter mounted on the CFHT. 
A careful analysis of the very young primary reveals that it has an approximately 
dipolar magnetic field geometry, strong chemical peculiarities, and strong 
surface chemical abundance inhomogeneities.  Thus the primary is very similar to an 
Ap/Bp star despite having completed less then 1.5\% of its main sequence life, 
and possibly still being on the pre-main sequence.  In contrast, a similar analysis
of the secondary reveals solar chemical abundances and no magnetic field.  
\end{abstract}

\section{Introduction}
\label{introduction}
Strong, globally-ordered magnetic fields have been known to exist in Ap/Bp stars for quite some time.  
However, the origin and evolution of these magnetic fields, as well as their impact on the observed peculiar 
chemical abundances, remains unclear.  Recently, 
magnetic fields have been discovered in some Herbig Ae and Be (HAeBe) stars 
(Donati et al., 1997; Hubrig et al., 2004; Wade et al., 2005; Wade et al., 2007).
HAeBe stars are pre-main sequence intermediate-mass stars and are the direct evolutionary progenitors of A and B stars.

The discovery of magnetic fields in HAeBe stars is of particular importance because it suggests a possible 
evolutionary link between Ap/Bp stars and magnetic HAeBe stars.  The observed magnetic fields in HAeBe stars 
appear to have similar strengths and geometries to those observed in Ap/Bp stars (Alecian et al., P27, these proceedings).
Furthermore, magnetic fields are detected in HAeBe stars with approximately the same frequency as in main sequence A and B stars, 5-10\% 
(Wade et al., 2007).

These recent discoveries prompted us to perform a detailed investigation of one particularly interesting case: 
the very young binary system HD 72106.

\section{HD 72106}
\label{HD 72106}
HD 72106 is a double star system with a 0.8\arcsec\, separation between the components.  
The brighter component (HD 72106A) was observed to possess a magnetic field by Wade et al. (2005), 
and the secondary was identified as being a HAeBe star by Vieira et al. (2003).  

The HD 72106 system was observed by Hipparcos, and solved as a binary, producing a ``good quality'' solution. 
Very large proper motions in both RA: $-5.18 \pm 1.08$ mas~yr$^{-1}$, 
and in DEC: $9.73 \pm 1.36$ mas~yr$^{-1}$  were observed, with no evidence for relative motion between the components (ESA, 1997).  
The Hipparcos parallax for the system places the two stars at a distance of $288^{+202}_{-84}$ pc. 
The radial velocities we determine from our observations of both components are identical at $22 \pm 1$ \kms.
Thus the stars are at approximately the same point in three dimensional space, and are moving together in three dimensions.  
This strongly suggesting that the system is in fact a true binary.  

Based on both Balmer line and metallic line fitting, we find \teff\, = $11000 \pm 1000$ K 
and \lgg\, = $4.0 \pm 0.5$ for the primary, and \teff\, = $8750 \pm 500$ K and \lgg\, = $4.0 \pm 0.5$ for the secondary.

\begin{figure}[htb]
\centering
\includegraphics[width=2.5in]{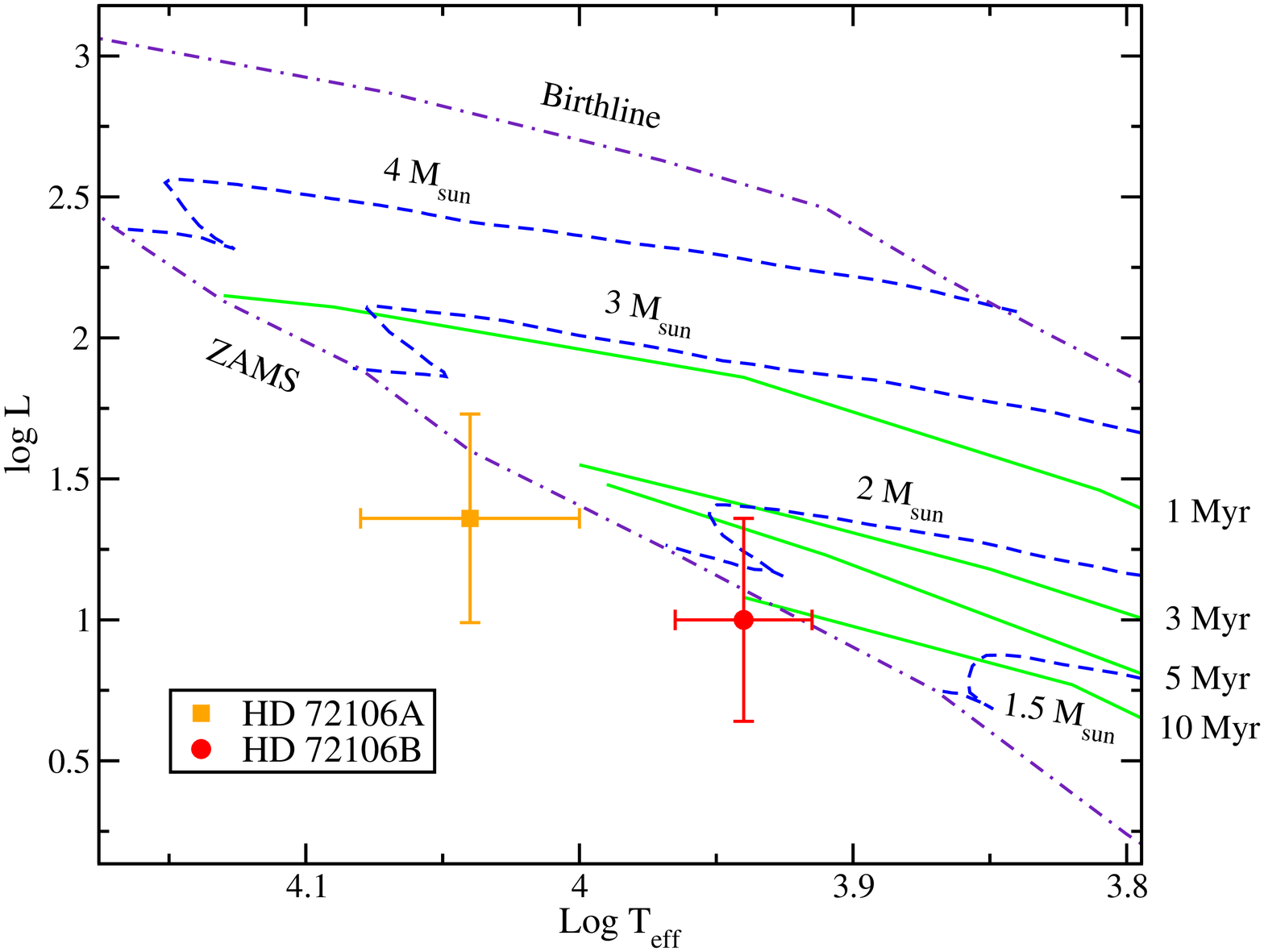}
\caption{The pre-main sequence H-R diagram of the HD~72106 system.  
Evolutionary track (solid lines) and isochrones (dashed lines), as well as 
the ZAMS and the birthline (for an accretion rate of $16^{-5}$ M$_{\odot}$ yr$^{-1}$) from Palla \& Stahler (1993) are shown.}
\label{h-r}
\end{figure}

The two stars can be placed on the H-R diagram, as seen in Fig.~\ref{h-r}.  
When compared with the evolutionary tracks of Palla \& Stahler (1993), we find that 
both stars sit near the zero-age main sequence (ZAMS).  Based on the position of the secondary 
(which we infer to be a pre-main sequence star based on emission at H$\alpha$) 
we can constrain the age of the system to be between 3 and 13 Myr.  
At the younger end of this range, the primary would still 
have $\sim \frac{1}{4}$ of its pre-main sequence lifetime remaining.  However, at the older 
limit, the primary would already be on the main sequence, with a fractional main 
sequence age ($\tau$) of 0.015.  Thus the evolutionary status of the primary is unclear, 
although the star is certainly very young.

In this investigation we used a set of 20 high-resolution Stokes $I$ and $V$ spectra obtained with ESPaDOnS, 
the high-resolution echelle spectropolarimeter mounted on the CFHT.

\section{Magnetic Fields}
We applied the Least Squares Deconvolution (LSD) technique (Donati et al., 1997) to our observations 
of both HD 72106A \& B. 
We find clear evidence for a magnetic field in nearly all our Stokes $V$ profiles of HD 72106A, 
with maximum longitudinal field of $385 \pm 45$~G,
but we see no evidence for a magnetic field in any of our observations of HD 72106B 
(maximum longitudinal field of $-51 \pm 55$~G).  

In order to determine the magnetic field geometry of HD 72106, we need to know the rotation period of the star.
We searched for periodicity in our longitudinal magnetic field observations using a modified Lomb-Scargle method.  
We also applied this technique to both our Stokes $I$ and $V$ LSD profiles themselves; searching each pixel in the line profile 
for periodic variability, and then averaging the results over all pixels.  
Additionally, the phasing of LSD profiles produced by candidate periods was investigated by eye,
looking for variability physically consistent with rotation.
The one period meeting all these criteria, $0.63995 \pm 0.00014$ days, was adopted as the rotation period of the star.

Using this rotation period, together with the inferred stellar radius and \vs, 
we determined the inclination of the rotation axis to the line-of-sight of HD 72106A to be $i = 23 \pm 11^\circ$.
We modeled the magnetic field of HD 72106A by fitting the observed Stokes $V$ LSD profiles with synthetic
profiles computed by ZEEMAN2 (Landstreet, 1988; Wade et al., 2001).  
The mean Stokes $I$ LSD profile was fit, then the phased set of Stokes $V$ LSD profiles was fit 
using a dipole field model.  
From this procedure we derive the dipole polar strength  $B_{\rm p} = 1300 \pm 100$ G and obliquity $\beta = 60 \pm 5^\circ$, 
with $3\sigma$ error bars.  
An analysis of the longitudinal field measurements shows that they are also consistent with this model.

\begin{figure}[tbh]
\centering
\includegraphics[width=2.5in]{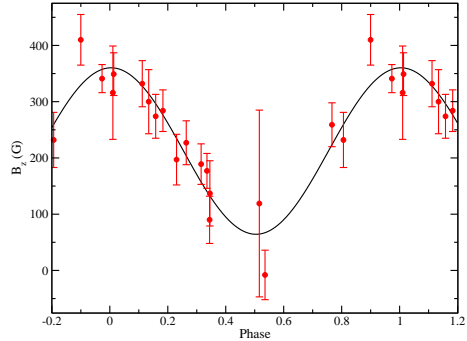}
\caption{Longitudinal magnetic field measurements, phased according to the rotation period of $0.63995 \pm 0.00014$ days,
together with the best fit sinusoid for the data.}
\label{phasedBz}
\end{figure}

\section{Abundance Analysis}
\label{Abundance Analysis}
\begin{figure}[htb]
\centering
\includegraphics[width=3.0in]{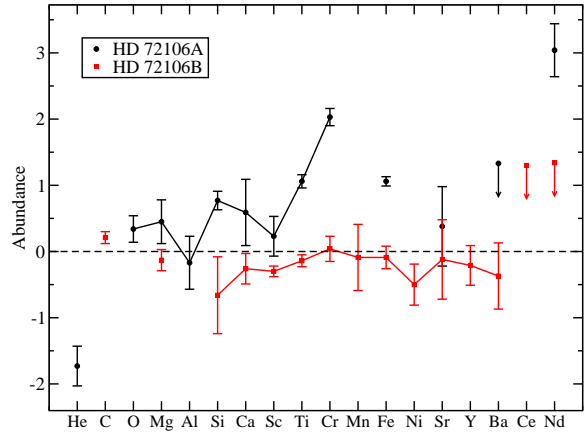}
\caption{Mean best fit abundances, relative to solar, for both HD 72106A \& B. Points marked with only an arrow indicate 
the value is an upper limit only.  Solar abundances are from Grevesse et al.(2005). }
\label{abun plot}
\end{figure}

The spectra of both components of HD 72106 were fit, using synthetic spectra calculated with 
the ZEEMAN2 spectrum synthesis code.  
Spectra were fit with a  Levenberg-Marquardt $\chi^2$ minimization method, 
and the results were extensively verified by eye.  
To approximate our magnetic observations, a 1 kG dipole magnetic field was included in the model of HD 72106A, and no field was used for HD 72106B.  
The final best fit values, reported in Fig.~\ref{abun plot}, are averages over seven spectral windows 
between 100 and 200 \AA\, in length, and the uncertainties are the standard deviations.  
When an abundance was based on only a few windows, a conservative estimate of the uncertainty was made by eye. 

For HD 72106A we find remarkably strong over-abundances of Cr, Fe and Nd.  Additionally, 
Si is overabundant by $\sim0.8$ dex while He appears to be underabundant by $\sim1.5$ dex.
We find \vs\, = 41.0 $\pm$ 0.3 \kms\, for this star.
This pattern of strong peculiarities is distinctive of Bp stars.

In contrast, for HD 72106B we find abundances that are consistent with solar, at least at the 2$\sigma$ level.  
Small departures from solar abundance for C and Sc may exist, however they 
are consistent with the normal scatter for an A star.
We find \vs\, = $53.9 \pm 1.0$ \kms\, and a microturbulence of $2.3 \pm 0.6$ \kms.

\section{Doppler Imaging}
We performed Doppler Imaging of HD 72106A using the INVERS12 code (Kochukhov et al., 2004).
LSD profiles were used for the Doppler Imaging, rather then individual lines, due to the S/N of our spectra.  
LSD profiles for specific elements were constructed using line masks containing only lines of one element.  

\begin{figure}[htb]
\centering
\includegraphics[width=3.5in]{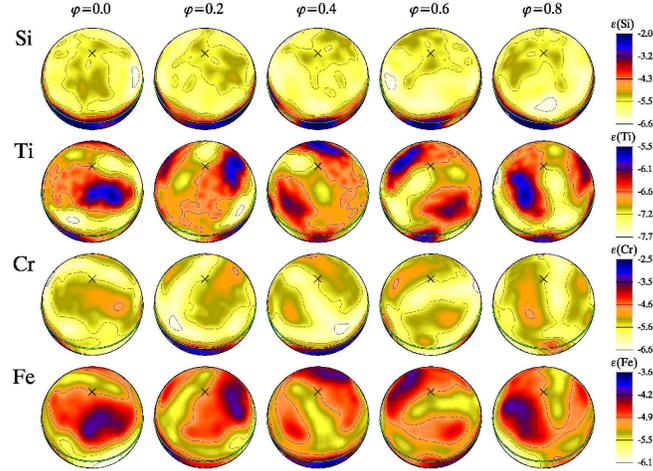}
\caption{Surface abundance maps of HD 72106A for Si, Ti, Cr and Fe.  }
\label{DI maps}
\end{figure}

Surface abundance maps were calculated for Si, Ti, Cr, and Fe, using a $30^\circ$ inclination angle, 
and are shown in Fig.~\ref{DI maps}.  
Strongly inhomogeneous distributions can be seen for all four elements.  Ti, Cr, and Si have similar distributions.  
The distribution of Ti appears to differ slightly from that of Cr and Fe in the southern hemisphere, 
but the inversion is poorly constrained in that region.  The large spot seen in the northern 
hemisphere near phase 0 in all four images is close to the location of the positive magnetic pole.

\section{Conclusions}
HD 72106A is in many respects very much like an Ap/Bp star.  It possess a predominantly dipolar magnetic field, 
strong chemical peculiarities, and strong surface abundance inhomogeneities.  
Despite this, the star is very young, having completed at most 1.5 \% of its main sequence life and possibly 
still being on the pre-main sequence.  In contrast, HD 72106B is chemically normal, 
and has no detectable magnetic field.  

The HD 72106 system provides evidence that chemical peculiarities can form at the ZAMS, 
and possibly even on the pre-main sequence.  Additionally, it provides support for an evolutionary 
link between magnetic HAeBe stars and Ap/Bp stars.  Finally this system raises questions about 
when in a star's evolution chemical peculiar first appear, in particular whether peculiarities 
can arise during the pre-main sequence or only early in the main sequence.

\acknowledgements
EA is funded by the Marie-Curie FP6 programme, while JDL
and GAW acknowledge support from NSERC and DND-ARP (Canada).


\end{document}